\documentclass[doublecol]{epl2} 
\usepackage{amsmath}
\title{Stability of superfluid phases in the 2D Spin-Polarized Attractive Hubbard Model}
\shorttitle{Stability of Superfluid Phases in the 2D Spin-Polarized Attractive Hubbard Model} %Insert here a short version of the title if it exceeds 70 characters

\author{A. Kujawa-Cichy\inst{1} \and R. Micnas\inst{1}}
\shortauthor{A. Kujawa-Cichy \etal}

\institute{                    
  \inst{1} Solid State Theory Division, Faculty of Physics, Adam Mickiewicz University, Umultowska 85,
61-614 Pozna\'n, Poland\\
}

\abstract{We study the evolution from the weak coupling (BCS-like limit) to the strong coupling limit of tightly bound local pairs (LP's) with increasing attraction, in the presence of the Zeeman magnetic field ($h$) for $d=2$, within the spin-polarized attractive Hubbard model. The broken symmetry Hartree approximation {as well as the} strong coupling expansion are used. We also apply the Kosterlitz-Thouless (KT) scenario to determine the phase coherence temperatures. For spin independent hopping integrals ($t^{\uparrow}=t^{\downarrow}$), we find no stable homogeneous polarized superfluid (SC$_M$) state in the ground state for the strong attraction  and obtain that for a two-component Fermi system on a 2D lattice with population imbalance, phase separation (PS) is favoured for a fixed particle concentration, even on the LP (BEC) side. We also examine the influence of spin dependent hopping integrals (mass imbalance) on the stability of the  SC$_M$ phase. We find a topological quantum phase 
transition (Lifshitz type) from the unpolarized superfluid phase (SC$_0$) to SC$_M$ and tricritical points in the ($h-|U|$) and ($t^{\uparrow} / t^{\downarrow} - |U|$) ground state phase diagrams. We also construct the finite temperature phase diagrams for both $t^{\uparrow} = t^{\downarrow}$ and $t^{\uparrow}\neq t^{\downarrow}$ and analyze the possibility of occurrence of a spin polarized KT superfluid.}

\begin{document}

\maketitle

\section{Introduction}

\begin{figure}[t!]%
\includegraphics*[width=.3\textwidth,height=\linewidth,angle=270]{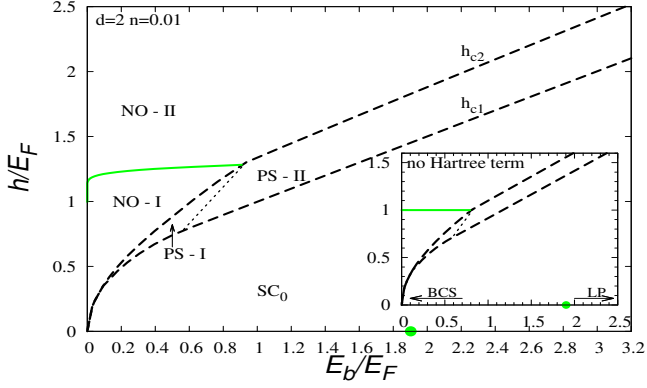}
\caption{Magnetic field vs. binding energy phase diagram both with and without the Hartree term (inset), in units of the lattice Fermi energy, at $T=0$ and fixed $n=0.01$ for $d=2$, $r=1$. $SC_0$ -- unpolarized superconducting state with $n_{\uparrow}=n_{\downarrow}$, {LP} -- {tightly bound local pairs}. Green solid line separates partially polarized (NO-I) and fully polarized (NO-II) normal states. PS-I ($SC_0$+NO-I) -- partially polarized phase separation, PS-II ($SC_0$+NO-II) -- fully polarized phase separation, $h_{c1}$, $h_{c2}$ -- critical fields defining the PS region. The green point shows the BCS-{LP} crossover point ($U/t=-4.01959$).
}
\label{fig1}
\end{figure}

Unconventional superconductivity with nontrivial Cooper pairing and spin-polarized superfluidity have been investigated in recent years.
The development of experimental techniques in cold atomic Fermi gases with tunable attractive interactions (through Feshbach resonance) has allowed the study the BCS-BEC crossover and the properties of exotic states in these systems \cite{ketterle}. 

The presence of a magnetic field ($h$), population imbalance or mass imbalance (spin dependent hopping integrals ($t^{\uparrow}\neq t^{\downarrow}$, $t^{\uparrow}/t^{\downarrow}\equiv r$ -- mass ratio) introduces a mismatch between Fermi surfaces (FS). This makes possible the formation of Cooper pairs across the spin-split Fermi surface with non-zero total momentum ($\vec{k} \uparrow$, $-\vec{k}+\vec{q} \downarrow$) (Fulde, Ferrell, Larkin and Ovchinnikov \cite{fulde} (FFLO) state). Another kind of pairing and phase coherence that can appear is the spatially homogeneous spin-polarized superfluidity (SC$_M$) (called breached pair (BP) state or Sarma phase \cite{sarma}), which has a gapless spectrum for the majority spin species \cite{Sheehy}. 

First theoretical studies of Fermi condensates in systems with spin and mass imbalances have shown that the BP state can have excess fermions with two FS's at $T=0$ (BP-2 or interior gap state) \cite{Wilczek,Wilczek-2,Iskin}. However, the problem of stability of the BP-2 state is still open. According to some investigations, the interior gap state proposed by Liu and Wilczek \cite{Wilczek} is always unstable even for large mass ratio $r$ and PS is favoured \cite{Parish}. 

At $T=0$, for strong attraction, the $SC_M$ phase occurs in the three-dimensional imbalanced Fermi gases \cite{Sheehy, Parish} as well as in the {3D} spin-polarized attractive Hubbard model in the dilute limit (for $r=1$ and $r\neq 1$ \cite{kujawa2}). The $SC_M$ phase is a specific superfluid state consisting of a coherent mixture of LP's (hard-core bosons)  and excess spin-up fermions (Bose-Fermi mixture). This state can only have one FS.

In this paper we study the superfluid phases in the attractive Hubbard model (AHM) ($U<0$) in a magnetic field with spin-dependent hopping:
\begin{equation}
\label{ham}
H=\sum_{ij\sigma} (t_{ij}^{\sigma}-\mu \delta_{ij})c_{i\sigma}^{\dag}c_{j\sigma}+U\sum_{i} n_{i\uparrow}n_{i\downarrow}-h\sum_{i}(n_{i\uparrow}-n_{i\downarrow}),
\end{equation}
where: $\sigma=\uparrow,\downarrow$, $n_{i\uparrow}=c_{i\uparrow}^{\dag}c_{i\uparrow}$, $n_{i\downarrow}=c_{i\downarrow}^{\dag}c_{i\downarrow}$, $t_{ij}^{\sigma}$ -- hopping integrals, $U$ -- on-site interaction, $\mu$ -- chemical potential. 
The gap parameter is defined by: $\Delta=-\frac{U}{N}\sum_i \langle c_{i \downarrow} c_{i \uparrow} \rangle =-\frac{U}{N}\sum_{\vec{k}} \langle c_{-\vec{k} \downarrow} c_{\vec{k} \uparrow} \rangle$. Applying the broken symmetry Hartree (BCS-Stoner) approximation, we obtain the grand canonical potential $\Omega$ and the free energy $F$ \cite{kujawa}. Using the free energy expression, one gets the equations for the gap: $\Delta=-\frac{U}{N}\sum_{\vec{k}} \frac{\Delta}{2\omega_{\vec{k}}}(1-f(E_{\vec{k}\uparrow})-f(E_{\vec{k}\downarrow}))$, particle number (which determines $\mu$): $n=n_{\uparrow}+n_{\downarrow}$, $n_{\sigma}=\frac{1}{N} \sum_{\vec{k}} \langle c_{\vec{k} \sigma}^{\dag} c_{\vec{k} \sigma} \rangle=\frac{1}{N}\sum_{\vec{k}}(|u_{\vec{k}}|^2 f(E_{\vec{k}\sigma})+|\nu_{\vec{k}}|^2 f(-E_{\vec{k}-\sigma}))$ and spin magnetization: $M=n_{\uparrow}-n_{\downarrow}$, where: $f(E_{\vec{k}\sigma})=1/(\exp(\beta E_{\vec{k}\sigma})+1)$, $\beta=1/k_B T$, $E_{\vec{k}\downarrow, \uparrow}= \pm (-t^{\downarrow}+t^{\
uparrow})\Theta_{\vec{k}}\pm \frac{UM}{2}\pm h+\omega_{\vec{k}}$, $\omega_{\vec{k}}=\sqrt{((-t^{\uparrow}-t^{\downarrow})\Theta_{\vec{k}}-\bar{\mu})^2+|\Delta|^2}$, $|\nu_{\vec{k}}|^2=\frac{1}{2} \Big(1-\frac{\xi_{\vec{k} \uparrow}+\xi_{\vec{k} \downarrow}}{2\omega_{\vec{k}}}\Big)$, $|u_{\vec{k}}|^2=1-|\nu_{\vec{k}}|^2$, $\xi_{\vec{k} \sigma}=\epsilon_{\vec{k} \sigma} -\bar{\mu}$, $\epsilon_{\vec{k} \sigma}=-2t^{\sigma}\Theta_{\vec{k}}$, $\Theta_{\vec{k}}=\sum_{l=1}^{d} \cos(k_l a_l)$ ($d=2$ for two-dimensional lattice), $a_l=1$ in further considerations, $\bar{\mu}=\mu-\frac{Un}{2}$. 
%$\frac{\partial F}{\partial \Delta}=0$, $\frac{\partial F}{\partial \mu}=0$, $M=-\frac{1}{N}\frac{\partial F}{\partial h}$, respectively. 
The equations take into account the spin polarization ($P=(n_{\uparrow}-n_{\downarrow})/n$) in the presence of a magnetic field and spin-dependent hopping ($t^{\uparrow}\neq t^{\downarrow}$, $\frac{t^{\uparrow}+t^{\downarrow}}{2}=t$, $t^{\uparrow}/t^{\downarrow}\equiv r$) {\cite{Wilczek-2,kujawa}}.

We also calculate the superfluid density $\rho_s(T)$ which for $t^{\uparrow} \neq t^{\downarrow}$ takes the form: %\eqref{longeq} (see below),
\begin{floatequation}
\mbox{\textit{see Eq.~\eqref{ro_s},}}
\end{floatequation}
\addtocounter{equation}{-1}
\begin{widetext}
\begin{eqnarray}
\label{ro_s}
\rho_s(T)&=&\frac{1}{4N}\sum_{\vec{k}} \Bigg\{ \frac{\partial ^{2}\epsilon^{+}_{\vec{k}}}{\partial k_x^2} -\frac{1}{2}\Bigg[\frac{\partial ^{2}\epsilon^{-}_{\vec{k}}}{\partial k_x^2}+\frac{\epsilon^{+}_{\vec{k}}}{\omega_{\vec{k}}}   \Bigg(\frac{\partial ^{2}\epsilon^{+}_{\vec{k}}}{\partial k_x^2}\Bigg) 
+\Bigg(\frac{\partial \epsilon^{-}_{\vec{k}}}{\partial k_x}\Bigg)^2 \frac{|\Delta|^2}{\omega_{\vec{k}}^3} \Bigg] \tanh \Bigg(\frac{\beta E_{\vec{k}\uparrow}}{2}\Bigg)\nonumber \\
&+&\frac{1}{2}\Bigg[\frac{\partial ^{2}\epsilon^{-}_{\vec{k}}}{\partial k_x^2}-\frac{\epsilon^{+}_{\vec{k}}}{\omega_{\vec{k}}} \Bigg(\frac{\partial ^{2}\epsilon^{+}_{\vec{k}}}{\partial k_x^2}\Bigg)  
-\Bigg(\frac{\partial \epsilon^{-}_{\vec{k}}}{\partial k_x}\Bigg)^2 \frac{|\Delta|^2}{\omega_{\vec{k}}^3} \Bigg] \tanh \Bigg(\frac{\beta E_{\vec{k}\downarrow}}{2}\Bigg)\nonumber \\
&+&\Bigg[\frac{\partial \epsilon^{+}_{\vec{k}}}{\partial k_x}+ \frac{\epsilon^{+}_{\vec{k}}}{\omega_{\vec{k}}} \Bigg(\frac{\partial \epsilon^{-}_{\vec{k}}}{\partial k_x}\Bigg)\Bigg]^2 
\frac{\partial f(E_{\vec{k}\uparrow})}{\partial E_{\vec{k}\uparrow}}+\Bigg[\frac{\partial \epsilon^{+}_{\vec{k}}}{\partial k_x}- \frac{\epsilon^{+}_{\vec{k}}}{\omega_{\vec{k}}} \Bigg(\frac{\partial \epsilon^{-}_{\vec{k}}}{\partial k_x}\Bigg)\Bigg]^2 \frac{\partial f(E_{\vec{k}\downarrow})}{\partial E_{\vec{k}\downarrow}}  \Bigg\},
\end{eqnarray}
\end{widetext}
where: $\epsilon_{\vec{k}}^{+}=\frac{\xi_{\vec{k}\uparrow}+\xi_{\vec{k} \downarrow}}{2}$, $\epsilon_{\vec{k}}^{-}=\frac{\xi_{\vec{k}\uparrow}-\xi_{\vec{k} \downarrow}}{2}-\bar{h}$, $\bar{h}=h+\frac{UM}{2}$.
%\begin{floatequation}
%\mbox{\textit{see Eq.~\eqref{longeq}}}
%\end{floatequation}

%\begin{eqnarray}
%\label{ro_s}
%&&\hspace{-0.5cm}\rho_s(T)=\frac{1}{4N}\sum_{\vec{k}} \Bigg\{ \frac{\partial ^{2}\epsilon^{+}_{\vec{k}}}{\partial k_x^2} -\frac{1}{2}\Bigg[\frac{\partial ^{2}\epsilon^{-}_{\vec{k}}}{\partial k_x^2}+\frac{\epsilon^{+}_{\vec{k}}}{\omega_{\vec{k}}}   \Bigg(\frac{\partial ^{2}\epsilon^{+}_{\vec{k}}}{\partial k_x^2}\Bigg) \nonumber \\ 
%&&\hspace{-0.5cm}+\Bigg(\frac{\partial \epsilon^{-}_{\vec{k}}}{\partial k_x}\Bigg)^2 \frac{|\Delta|^2}{\omega_{\vec{k}}^3} \Bigg] \tanh \Bigg(\frac{\beta E_{\vec{k}\uparrow}}{2}\Bigg)+\frac{1}{2}\Bigg[\frac{\partial ^{2}\epsilon^{-}_{\vec{k}}}{\partial k_x^2}-\frac{\epsilon^{+}_{\vec{k}}}{\omega_{\vec{k}}} \Bigg(\frac{\partial ^{2}\epsilon^{+}_{\vec{k}}}{\partial k_x^2}\Bigg) \nonumber \\ 
%&&\hspace{-0.4cm}-\Bigg(\frac{\partial \epsilon^{-}_{\vec{k}}}{\partial k_x}\Bigg)^2 \frac{|\Delta|^2}{\omega_{\vec{k}}^3} \Bigg] \tanh \Bigg(\frac{\beta E_{\vec{k}\downarrow}}{2}\Bigg)+\Bigg[\frac{\partial \epsilon^{+}_{\vec{k}}}{\partial k_x}+ \frac{\epsilon^{+}_{\vec{k}}}{\omega_{\vec{k}}} \Bigg(\frac{\partial \epsilon^{-}_{\vec{k}}}{\partial k_x}\Bigg)\Bigg]^2\nonumber \\  
%&&\hspace{-0.5cm}\times \frac{\partial f(E_{\vec{k}\uparrow})}{\partial E_{\vec{k}\uparrow}}+\Bigg[\frac{\partial \epsilon^{+}_{\vec{k}}}{\partial k_x}- \frac{\epsilon^{+}_{\vec{k}}}{\omega_{\vec{k}}} \Bigg(\frac{\partial \epsilon^{-}_{\vec{k}}}{\partial k_x}\Bigg)\Bigg]^2 \frac{\partial f(E_{\vec{k}\downarrow})}{\partial E_{\vec{k}\downarrow}}  \Bigg\},
%\end{eqnarray}

For $d=2$, $h=0$, the SC-NO transition in the AHM is of the KT type if $n\neq 1$, mediated by unbinding of vortices. The KT temperature ($T_c^{KT}$) can be determined from the universal relation:
\begin{equation}
\label{KT}
 k_B T_c^{KT}=\frac{\pi}{2} \rho_s (T_c^{KT}).
\end{equation}

In the strong coupling limit ($|U| \gg {t^{\uparrow}}, {t^{\downarrow}}$), AHM ($U<0$, $h=0$, $r\neq 1$) is mapped (via the canonical transformation {\cite{Robak,MicnasModern,Cazalilla}}) onto the following pseudo-spin model:

\begin{eqnarray}
\label{pseudospin}
&&H=-{\frac{1}{2}}\sideset{}{'}{\sum}_{i, j} J_{ij} (\rho_i^+ \rho_j^- +h.c.) +\sideset{}{'}{\sum}_{i, j} K_{ij} \rho_i^z \rho_j^z \nonumber \\ 
&&-{\tilde{\mu}} \sum_i (2\rho_i^z+1)-\frac{N}{4}J_0,
\end{eqnarray}
{and}: $n=\frac{1}{N} \sum_{i} \langle 2\rho_i^z+1\rangle$. $J_0=\sum_{j} J_{ij}$, $J_{ij}=2\frac{t_{ij}^{\uparrow} t_{ij}^{\downarrow}}{|U|}$, $K_{ij}=2\frac{(t_{ij}^\uparrow)^2+(t_{ij}^{\downarrow})^2}{2|U|}$, {$\tilde\mu=\mu+\frac{|U|}{2}$}, the pseudo-spin operators are: $\rho_i^+=c_{i\uparrow}^{\dag}c_{i\downarrow}^{\dag}$, {$\rho_i^-=c_{i\downarrow}c_{i\uparrow}$}, $\rho_i^z=\frac{1}{2}(n_{\uparrow}+n_{\downarrow}-1)$, primed sum excludes terms with $i=j$. The pseudo-spin operators satisfy the commutation rules of $s=\frac{1}{2}$ operators. It is worth noting that Hamiltonian (\ref{pseudospin}) operates in the subspace of states without the single occupancy. After an appropriate transformation to the bosonic operators: $\rho_i^{+}=b_i^{\dag}$,  $\rho_i^{-}=b_i$, $\rho_i^{z}=-\frac{1}{2}+b_i^{\dag}b_i$, the above Hamiltonian (\ref{pseudospin}) describes a system of hard-core bosons on a lattice with the commutation relations {\cite{Strong, MicnasModern}}: $[b_i^{\dag},
b_j]=(2n_i-1)\delta_{ij}$, $b_i^{\dag}b_i + b_ib_i^{\dag}=1$, where $n_i=b_i^{\dag}b_i$.

With the mass imbalance, it is possible that the charge density wave ordered (CO) state can develop for any particle concentration. The SC to CO is a first order transition at $h=0$, {$t^{\uparrow}\neq t^{\downarrow}$} and $n \neq 1$. The critical $n$ ($n_c$) (within the mean field approximation) {above} which superconductivity can coexist with commensurate CO is given by {\cite{MicnasModern,Dao}}):
$|n_c-1|=\sqrt{\frac{K_{{0}}-J_{{0}}}{K_{{0}}+J_{{0}}}}$.  %$|n_c-1|=\sqrt{(K_{{0}}-J_{{0}})/(K_{{0}}+J_{{0}}})$.
%\begin{equation}
%\label{nc}
%|n_c-1|=\sqrt{(K_{{0}}-J_{{0}})/(K_{{0}}+J_{{0}}}).
% |n_c-1|=\sqrt{\frac{K_{{0}}-J_{{0}}}{K_{{0}}+J_{{0}}}}.
%\end{equation}
Substituting expressions for $J$ and $K$, one obtains:
\begin{equation}
 n_c=1\pm \Big| \frac{r-1}{r+1} \Big|.
\end{equation}
{Away from half filling the quantum fluctuations can extend the region of stability of the SC phase at $T=0$ and enhance $n_c$ \cite{Strong,Yunoki}.} 
%{Away from half filling the quantum fluctuations at $T=0$ can enhance $n_c$ \cite{MicnasModern}.}
In further considerations we fix $n<n_c$ and consider mostly low n.

\begin{figure*}[t!]%
{%
\includegraphics*[width=.265\textwidth,height=5.9cm,angle=270]{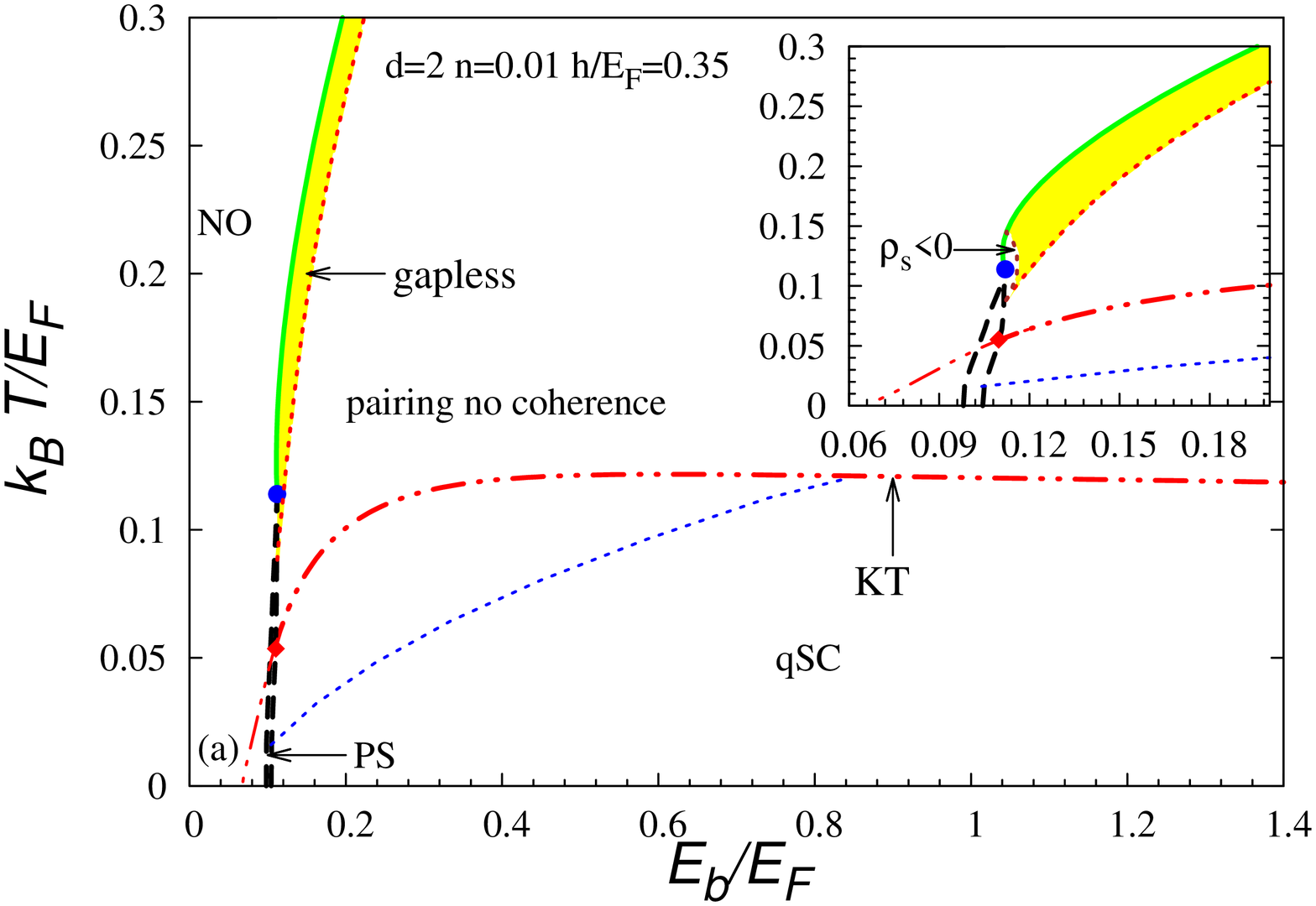}}\hfill
{%
\includegraphics*[width=.265\textwidth,height=5.9cm,angle=270]{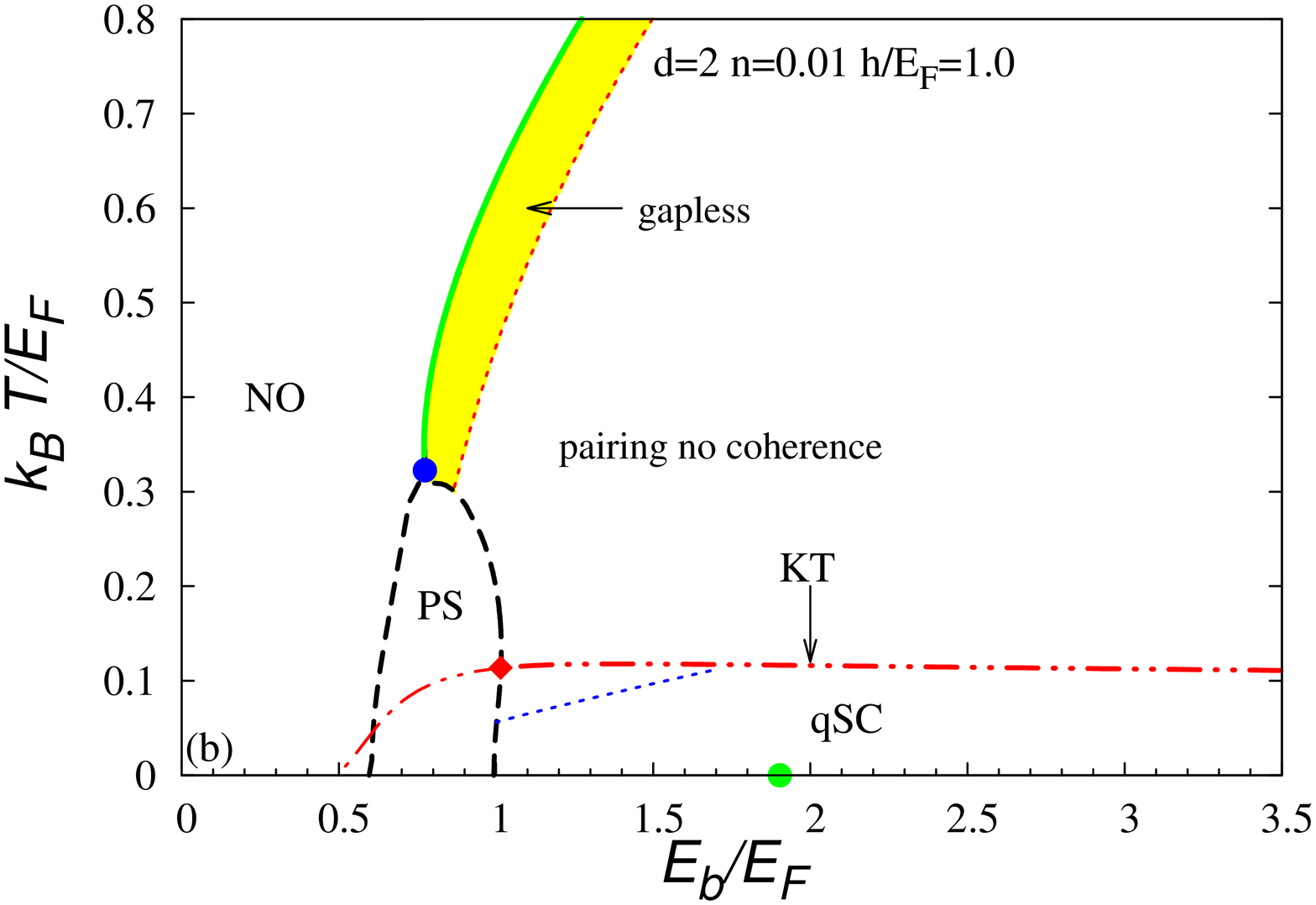}}\hfill
{%
\includegraphics*[width=.25\textwidth,height=5.9cm,angle=270]{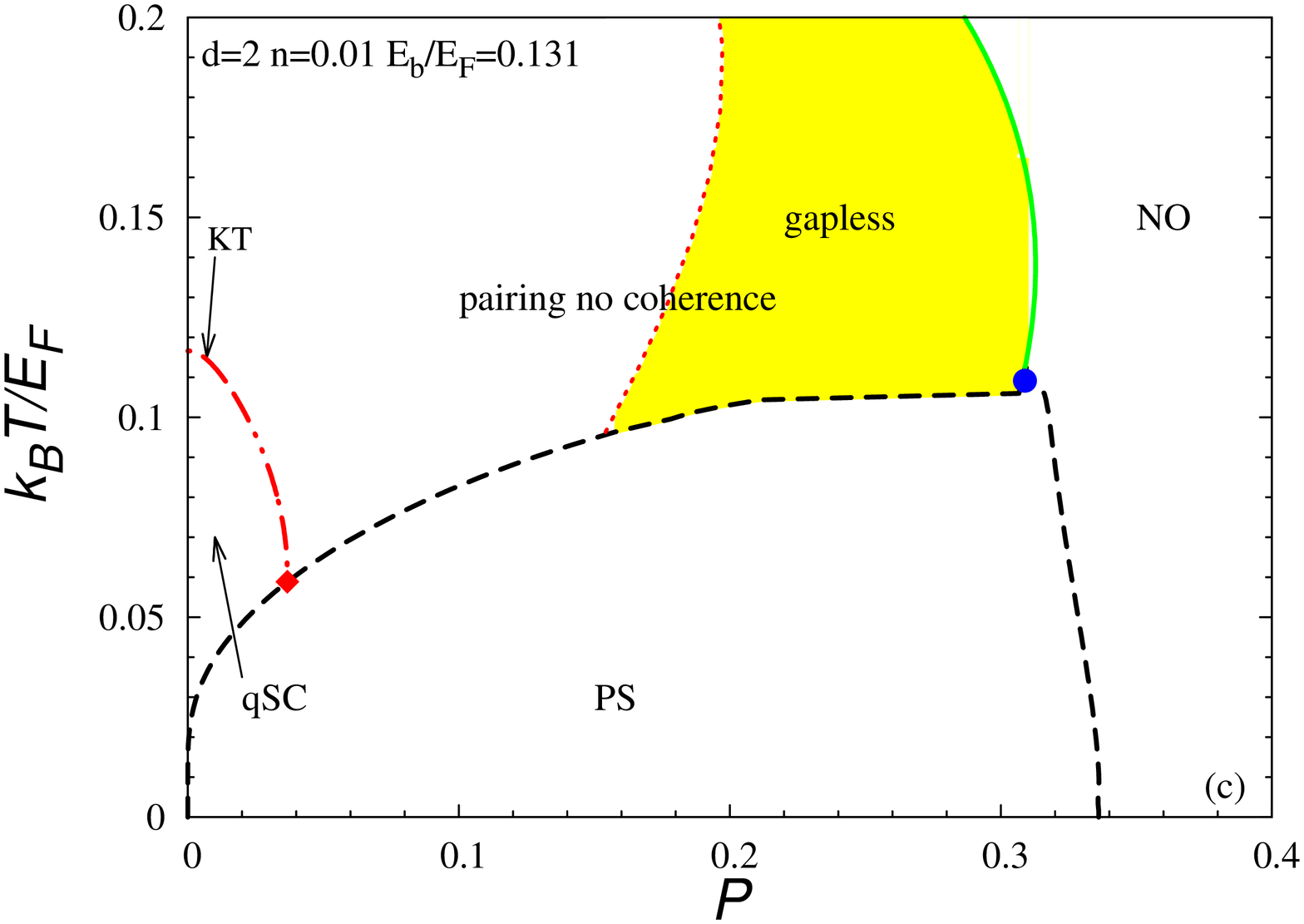}}%
%\includegraphics*[width=.34\textwidth,angle=270]{pole_h0022_n001_kT-dos-diag-Hartree-Eb-wstawka.ps}
%{\includegraphics*[width=.34\textwidth,angle=270]{pole_h0062_n001_kT-dos-diag-Hartree-Eb.ps}}
\caption{Temperature vs. binding energy phase diagrams in units of the lattice Fermi energy at (a) $h/E_F=0.35$ (inset -- details of a region around TCP), (b) $h/E_F=1$ and (c) T vs. P phase diagram at $E_b/E_F=0.131$.  Thick dashed-double dotted line (red color) is the KT transition line. Thin dash-double dotted line is the KT transition line to the metastable superfluid state. Thick solid line denotes transition from pairing without coherence region to NO within the Hartree approximation {with distinguished gapless region (yellow color)}. $qSC$ -- 2D quasi superconductor, PS -- phase separation. Below thin dotted line (blue color) -- $P<10^{-4}$. The green point (Fig. (b)) shows the BCS-{LP} crossover point {at $T=0$}; blue point is the MF TCP point.
}
\label{fig2}
\end{figure*}

\section{Results}

\begin{figure*}[]%
{%
\includegraphics*[width=.26\textwidth,height=5.9cm,angle=270]{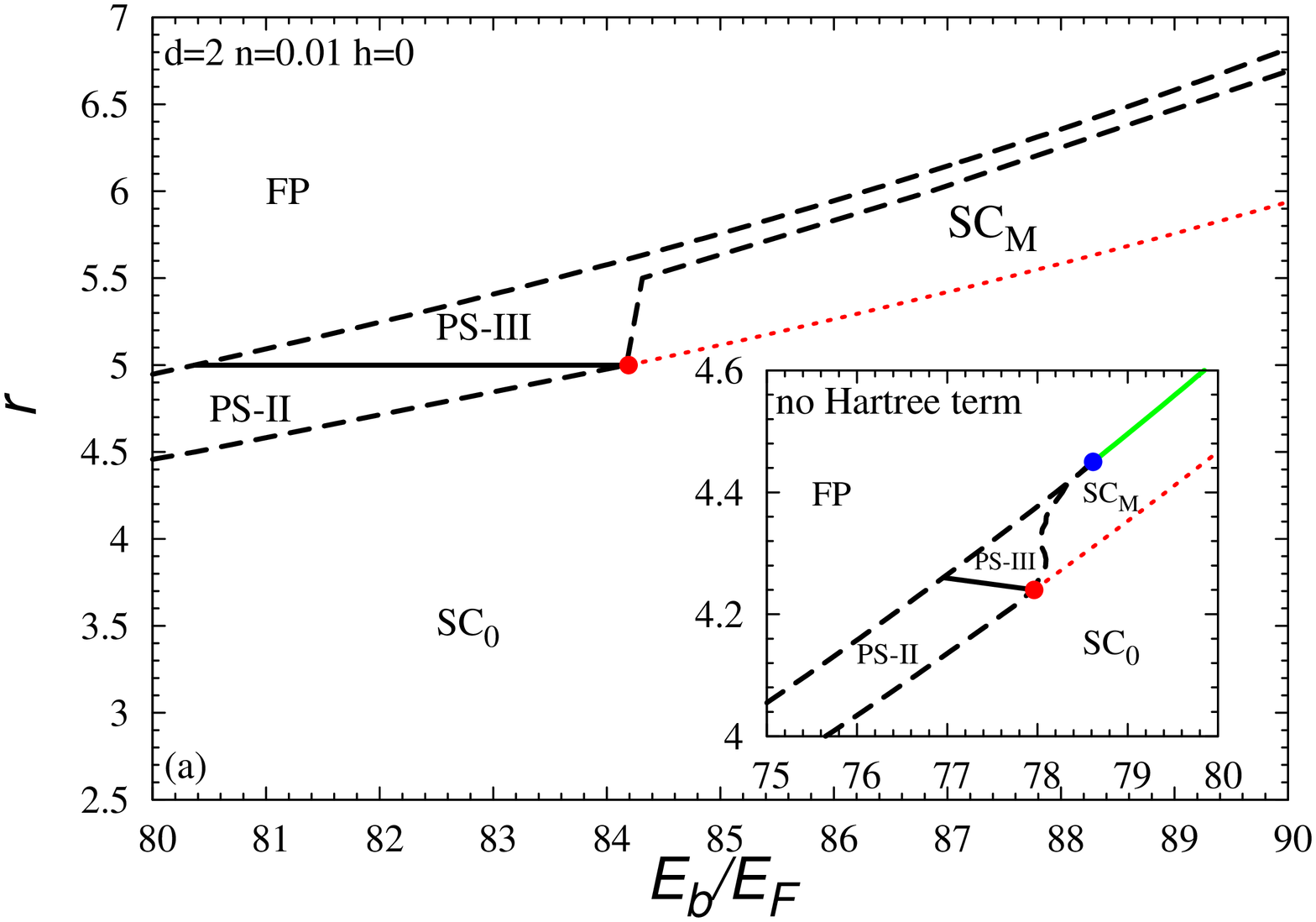}}\hfill
{%
\includegraphics*[width=.25\textwidth,height=5.9cm,angle=270]{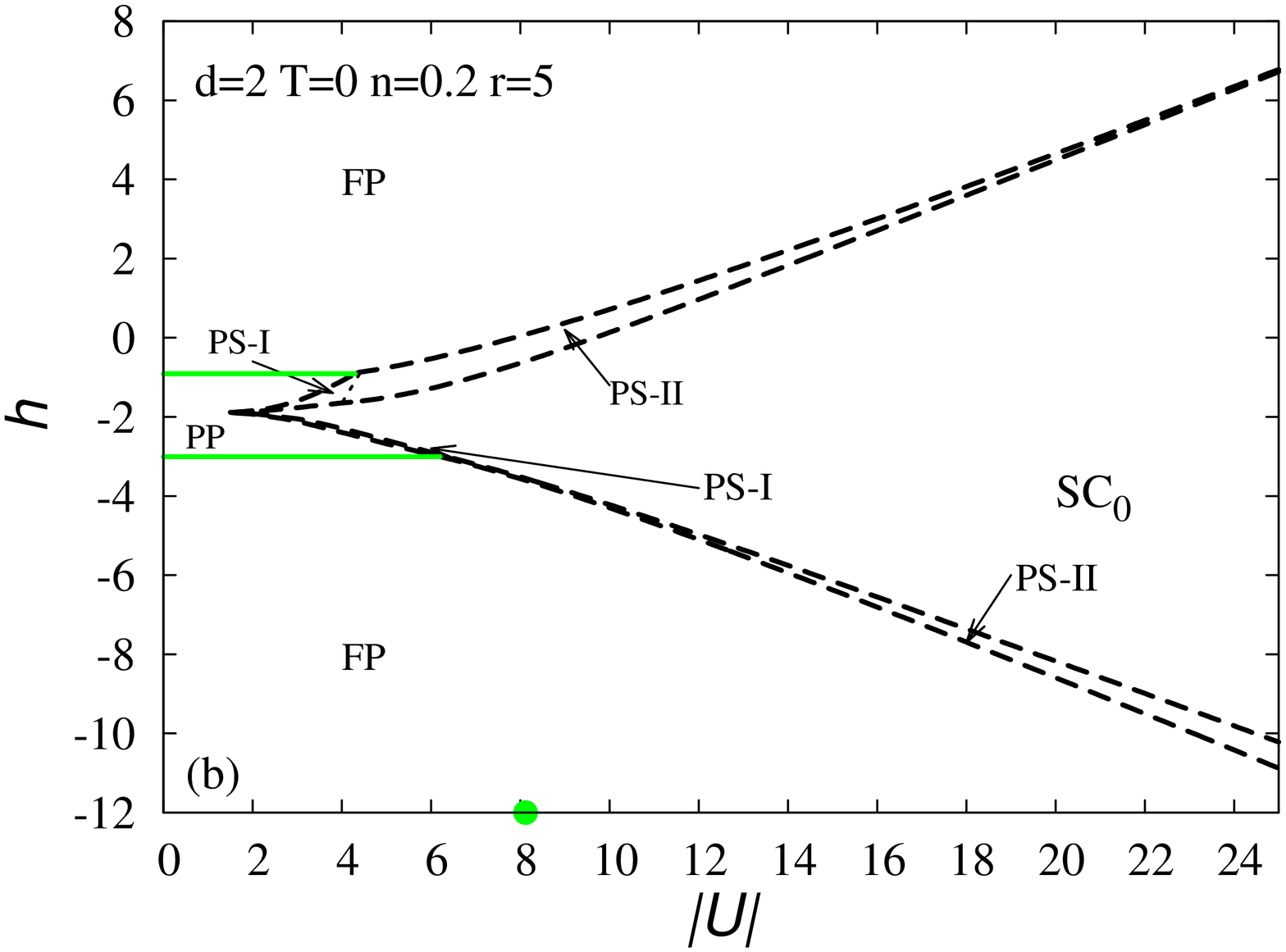}}\hfill
{%
\includegraphics*[width=.25\textwidth,height=5.9cm,angle=270]{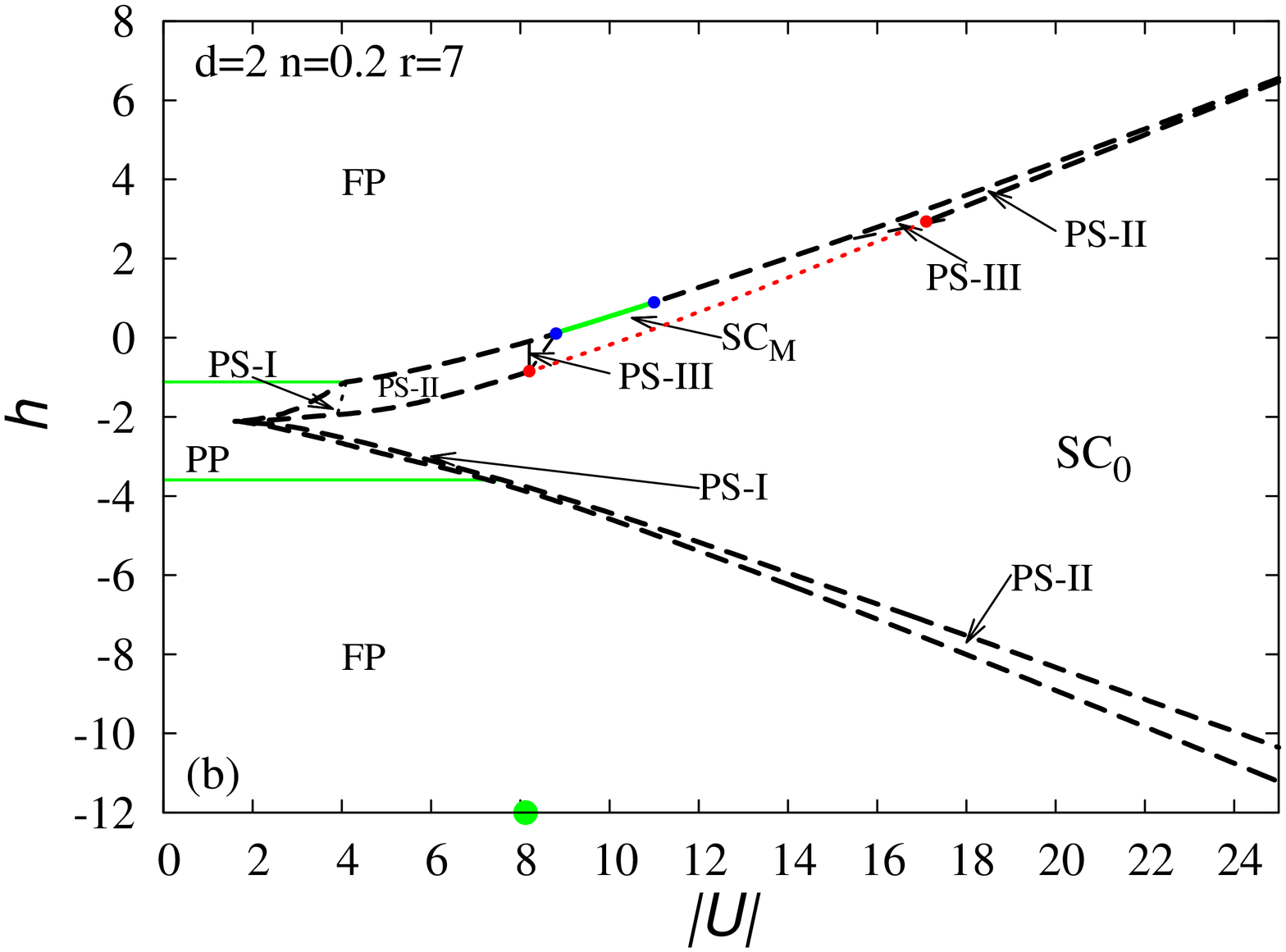}}%
%{\label{figs4c}%
%\includegraphics*[width=.2\textwidth,height=4.2cm,angle=270]{pole_U_n02_kT0-dos-h-t1_1t2_15-2D.ps}}%
\caption{%
Ground state phase diagrams for the $d=2$ square lattice: (a) mass imbalance vs. $E_b/E_F$ ({where $E_F$ is the lattice Fermi energy of unpolarized, non-interacting fermions with hopping $t$}) for $n=0.01$ and $h=0$ with and without Hartree term (inset). (b), (c) magnetic field vs. attractive interaction for two values of $r$ and $n=0.2$, without Hartree term. $SC_M$ -- magnetized superconducting state, PP -- partially polarized state, FP -- fully polarized state, PS-I -- (SC$_0$+PP), PS-II -- (SC$_0$+FP), PS-III -- ($SC_M$+FP). Red points -- $h_{c}^{SC_M}$, blue points  -- tricritical points, green points -- the BCS-{LP} crossover points in the SC$_0$ phase ($r=1$). The dotted red and the solid green lines are the $2^{nd}$ order transition lines.}
\label{fig3}
\end{figure*}

\begin{figure*}[t!]%
{%
\includegraphics*[width=.25\textwidth,height=5.9cm,angle=270]{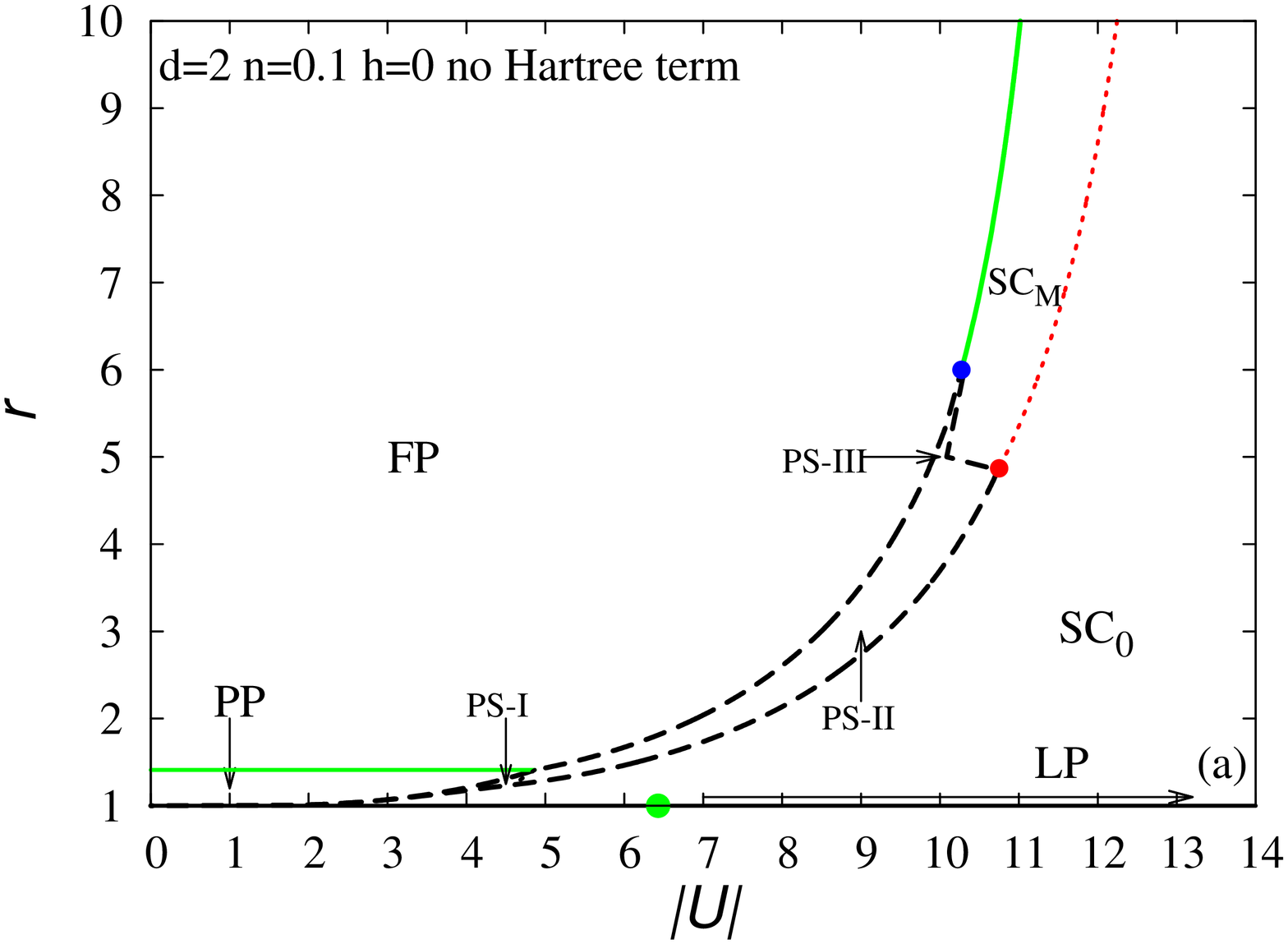}}\hfill
{%
\includegraphics*[width=.25\textwidth,height=5.9cm,angle=270]{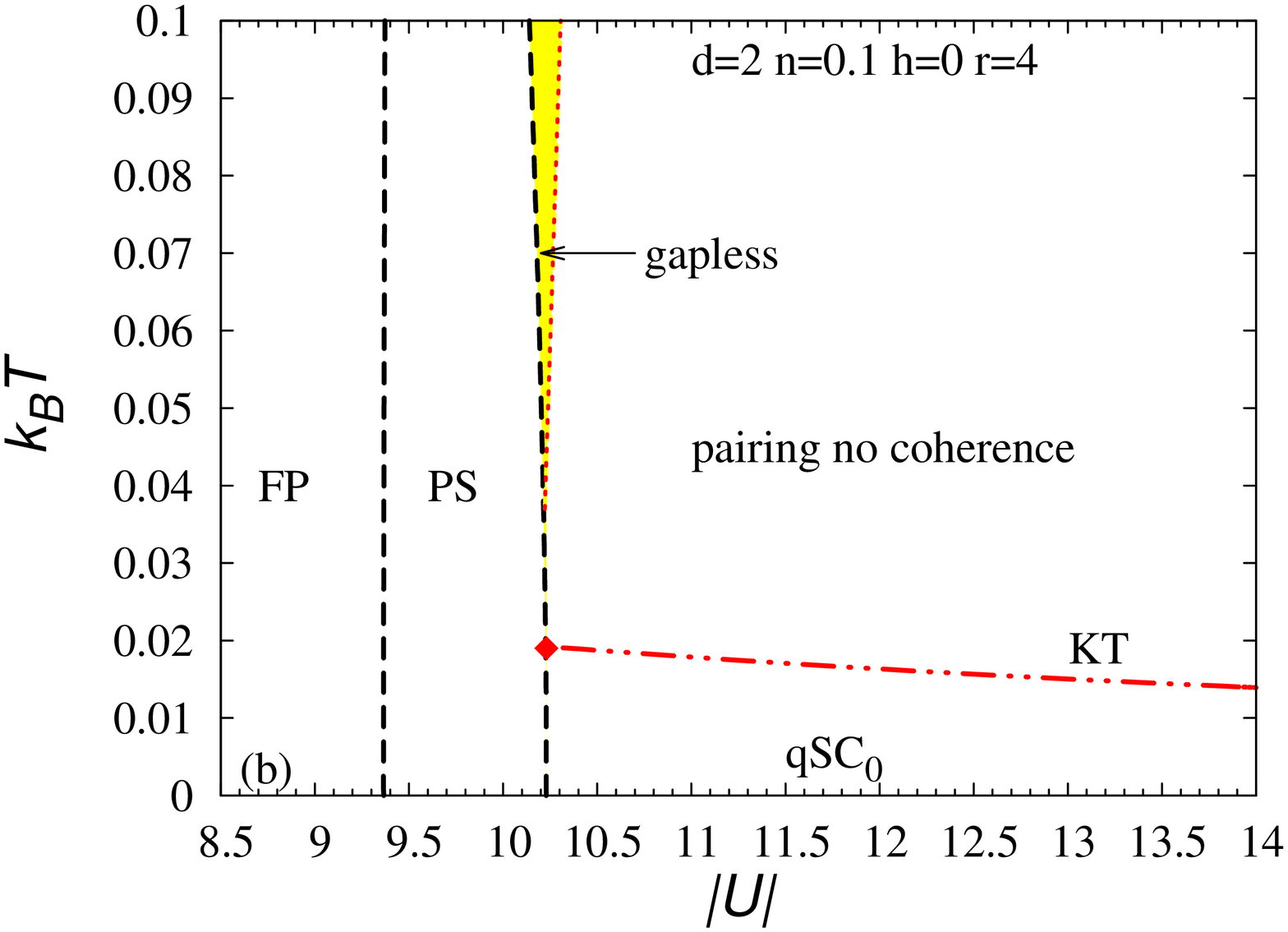}}\hfill
{%
\includegraphics*[width=.25\textwidth,height=5.9cm,angle=270]{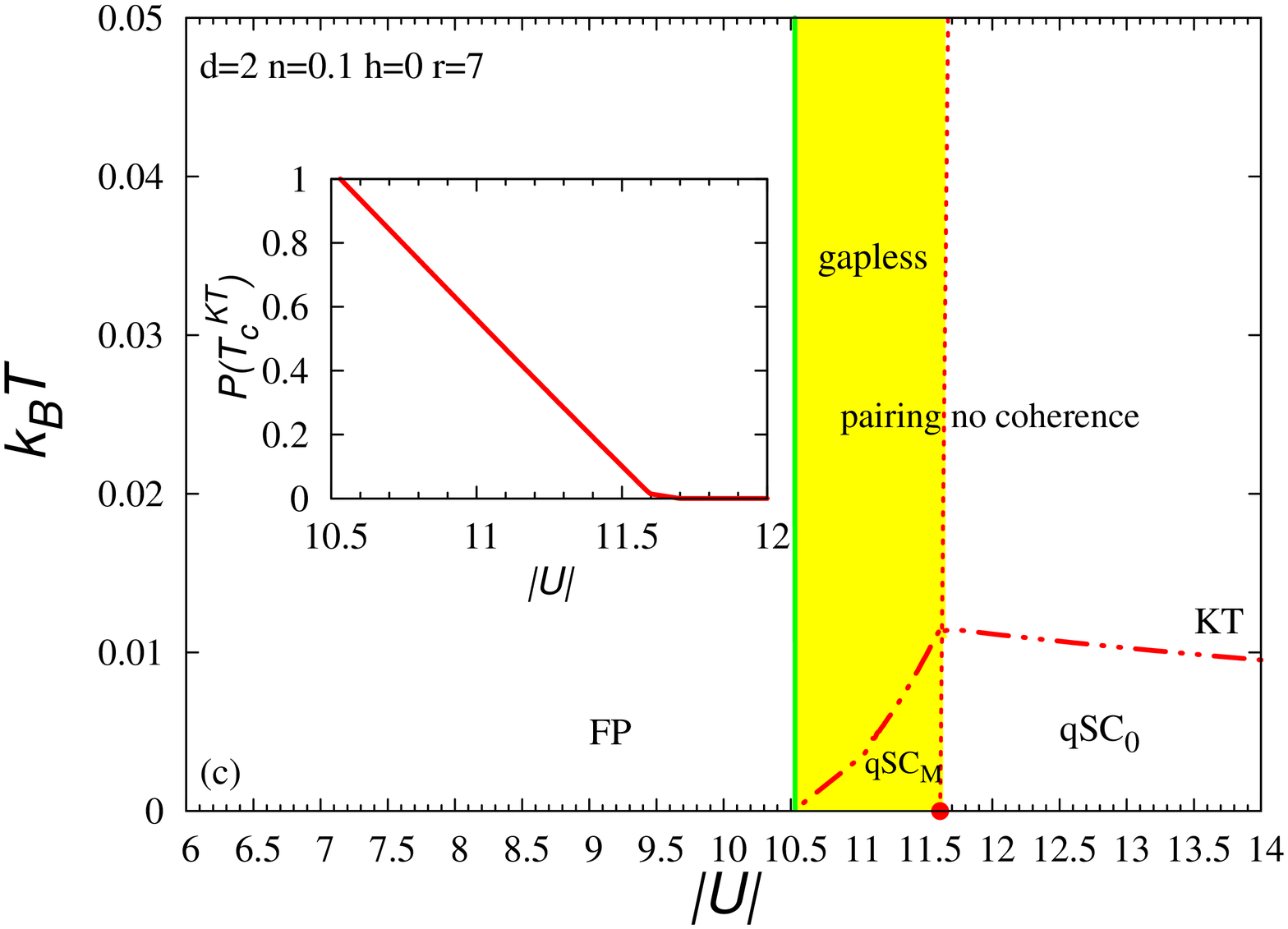}}%
%{\label{figs4c}%
%\includegraphics*[width=.2\textwidth,height=4.2cm,angle=270]{pole_U_n02_kT0-dos-h-t1_1t2_15-2D.ps}}%
\caption{%
(a) Ground state phase diagram $r$ vs. $|U|$ for $n=0.1$, $h=0$, Figs. (b-c) temperature vs. $|U|$ phase diagrams, at fixed $n=0.1$, $h=0$ for the square lattice. (b) $r=4$, (c) $r=7$ (inset -- $P\big(T_{c}^{KT}\big)$ vs. $|U|$). The thick dashed-double dotted line (red color) is the KT transition line.  Thick solid line denotes transition from pairing without coherence region to FP within the BCS approximation, dotted line (red color) is the second order transition line from qSC$_0$ to qSC$_M$ state at $T=0$ or from qSC$_0$ to the gapless region (yellow color) at $T\neq 0$. PS -- phase separation, qSC$_0$ -- 2D qSC without polarization, qSC$_M$ -- 2D qSC in the presence of polarization (a spin polarized KT superfluid). Red point at $T=0$ (see Fig. (c)) -- QCP for Lifshitz transition.}
\label{fig4}
\end{figure*}

We have performed an analysis of the evolution of superconducting properties from the weak to the strong coupling limit with increasing $|U|$, within the spin-polarized AHM. The system of self-consistent equations has been solved numerically for the 2D square lattice. First, the chemical potential has been fixed. The first order transition lines were determined from the condition $\Omega^{SC} =\Omega^{NO}$, where $\Omega^{SC}$, $\Omega^{NO}$ is the grand canonical potential of superconducting and normal state, respectively. Then, these results have been mapped onto the case of fixed $n$. The stability of all phases has been examined very thoroughly. In the following, we use $t$ as the unit.

According to the Leggett criterion \cite{leggett}, the BCS-BEC crossover takes place when the chemical potential $\bar{\mu}$ reaches the lower band edge. In the limiting case of two fermions in an empty lattice, one can perform an exact analysis. For the $s$-wave pairing, the two-particle binding energy ($E_b$) is given by \cite{MicnasModern}:
\begin{equation}
\frac{2D}{U}=-\frac{1}{N} \sum_{\vec{k}} \frac{1}{(\frac{E_b}{2D}+1)+\frac{\epsilon_{\vec{k}}}{D}},
\end{equation}
where: $D=zt$, $z=2d$ is the coordination number, $\epsilon_{\vec{k}} = -2t\Theta_{\vec{k}}$.
Since in $d=2$ the two-body bound state is formed for any attraction, one can express the pairing potential $|U|$ by $E_b$. This is of interest as far as a comparison with the continuum model of a dilute gas of fermions is concerned.
\\
%\subsection{$r=1$, $h\neq 0$, $T=0$}
\subsection{\textbf{(i)} $\boldsymbol{r=1}$, $\boldsymbol{h\neq 0}$, $\boldsymbol{T=0}$} 
In Fig. \ref{fig1} we present the ground state diagrams for arbitrary values of on-site attraction $|U|$ ($E_b$). The particle concentration is set at $n=0.01$. For fixed n, the 1$^{st}$ order SC-NO transition line in the ($\mu -h$) plane is replaced by PS region bounded by the two critical Zeeman fields $h_{c1}$ and $h_{c2}$. We find an unmagnetized SC$_0$ phase in the strong attraction (LP) limit. With increasing magnetic field, PS is energetically favoured i.e. even in the strong attraction limit, the Sarma or breached (BP-1) phase is unstable. Superconductivity is destroyed by pair breaking in the weak coupling regime. On the other hand, in the strong coupling regime, the transition from the superconducting to the normal state goes in addition through the phase separation (SC$_0$ $\rightarrow$ PS-II $\rightarrow$ NO-II). Similar behaviour persists for higher $n$. The phase diagram without the Hartree term has also been constructed and shown in the inset of Fig. \ref{fig1}. Let us point out that this 
diagram is in good agreement with the results for the continuum fermion model in $d=2$ \cite{He}. However, if we consider the model in continuum, the chemical potential changes its sign exactly at $E_b=2E_F$, which indicates the point of the BCS-BEC crossover. Our analysis gives a smaller ratio $E_b/E_F$. The differences are attributed to lattice effects, because the electron concentration is relatively small, but still finite. On the other hand, our results do not agree with the analysis in Ref. \cite{Du}, according to which the BP-2 state is stable at $r=1$ in $d=2$. 

If one takes corrections beyond mean-field (MF) into account, the existence of the spin imbalanced superfluid mixture of bosonic molecules and Fermi atoms can not be excluded in the BEC limit \cite{Tempere2}. We should also add that the deep BEC {side} is better described by a Boson-Fermion mixture of {hard-core bosons} and spin-up fermions. 
%however this requires a suitable generalization of the strong coupling expansion to allow single occupancies.
%(Eq. (\ref{pseudospin}), $t^{\uparrow}=t^{\downarrow}$)

\subsection{\textbf{(ii)} $\boldsymbol{r=1}$, $\boldsymbol{h\neq 0}$, $\boldsymbol{T\neq 0}$} Next, the results concerning the influence of the magnetic field on superfluidity at finite temperatures are presented. For $d=2$ AHM at $h=0$, the SC-NO transition is of the KT type i.e. below $T^{KT}$ a system has a quasi-long-range (algebraic) order which is characterized by a power law decay of the order parameter correlation function and non-zero superfluid stiffness. According to Eq. (\ref{KT}), the KT transition temperature is found from the intersection point of the straight line $\frac{2}{\pi} k_B T$ with the curve $\rho_s(T)$. In such a way, we can estimate the phase coherence temperatures and extend {the analysis of crossover from weak to strong coupling} to finite {$T$} \cite{tempere}.

Fig. \ref{fig2} shows ($T-E_b$) phase diagrams in units of Fermi energy, for fixed values of magnetic field: (a) $h/E_F=0.35$ (on the BCS side), (b) $h/E_F=1$ (in the intermediate couplings) and (c) the ($T-P$) diagram at fixed $E_b/E_F=0.131$. The solid lines (2$^{nd}$ order transition lines) and PS regions are obtained within the Hartree approximation. The thick dash-double dotted line (red color) denotes the KT transition determined from Eqs. (\ref{ro_s}-\ref{KT}). The system is a quasi superconductor (qSC) below $T_c^{KT}$. The polarization can be induced by thermal excitations of quasiparticles. Above $T_c^{KT}$ but below $T_c^{HF}$ (pair breaking temperature), pairs exist but without a long-range phase coherence. In this region one has a pseudogap behavior. The temperatures $T_c^{KT}$ are generally much smaller than $T_c^{HF}$, but reducing the attraction, in the absence of magnetic field, the difference between $T_c^{KT}$ and $T_c^{HF}$ decreases in a weak coupling limit. At $h=0$ and $E_b \ll 1$, $T_
c^{HF}\neq 0$ and $T_c^{KT}\neq 0$. When the magnetic field increases, $T_c^{HF}=0$, below a definite value of the binding energy. This critical $E_b$ increases with $h$ (Fig. \ref{fig1}). In the strict BCS-MFA diagram the tricritical point (TCP) exists at finite magnetic fields. In this mean-field TCP, the SC MF phase, the NO state and PS coexist. There is also {''TCP''} on the KT curve in which three states meet: the qSC phase, the state of incoherent pairs and PS (Fig. \ref{fig2}(a)-(b)). The PS range widens with increasing $h$ and the distance between the TCP MF and KT {''TCP''} is larger. As shown in Fig. \ref{fig2}(c), the effect of finite P on the KT superfluid state is strong. For $r=1$, the KT phase is restricted to the weak coupling region and low values of P. Increasing polarization favours phase of incoherent pairs. The range of occurrence of $qSC$ in the presence of P widens in the weak coupling regime with increasing $n$. In turn, the $qSC$ state is highly reduced with 
increasing attractive interaction even for low population imbalance. 

By the analysis of the quasiparticle excitation spectrum, we also find a gapless region (yellow color in diagrams), for $h> \Delta$ (the BCS side) and for $h>E_{g}/2$, where $E_g=2\sqrt{(\bar{\mu}-\epsilon_0)^2+|\Delta |^2}$ (on the {LP} side). If $r=1$ this gapless region can only be realized at $T>0$ and has excess fermions with two FS in the weak coupling limit. {In Fig. (\ref{fig2}) the gapless region is distinguished within the state of incoherent pairs, i.e. is nonsuperfluid.} In the strong coupling regime, the temperature can induce the spin-polarized gapless region (in a state of incoherent pairs) with one FS. This is in contrast to the 3D case where the BP-1 phase can be stable even without mass imbalance at $T=0$ and low $n$ \cite{kujawa2}. In the strong coupling limit, $T_c^{KT}$ does not depend on magnetic field and approximately approaches $k_BT_c^{KT}/E_F \approx \frac{t}{2|U|} (1-\frac{n}{2})$ for $|U|\gg t$, $E_F =2\pi tn$, in contrast to the continuum model which yields in 
that limit $k_B T_{c}^{KT}/E_F=\frac{1}{8}$ \cite{tempere}. For $k_B T<<|U|$, there exist only LP's not broken by the magnetic field and the system is equivalent to that of hard-core Bose gas on a lattice. The thin dash-double dotted line in Fig. \ref{fig2}(a-b) inside the NO state marks the region where $\Omega$ has two minima (below the curve): lower at $\Delta =0$ and higher at $\Delta \neq 0$. It means that there can exist a metastable superconducting state.

An interesting aspect of the analysis is the influence of the Hartree term on the phase diagrams. First, the presence of the Hartree term leads to the reentrant transition (RT) in the weak coupling limit (Fig. \ref{fig2}(a)), which is not observed in the phase diagrams without the Hartree term. We also find a narrow region around MF TCP in which $\rho_s<0$ although $\Omega^{SC}< \Omega^{NO}$ (Fig. \ref{fig2}(a) inset), in the phase diagram on the BCS side with the Hartree term. If RT exists, it becomes dynamically unstable because $\rho_s<0$. In addition, the Hartree term causes an increase in the Chandrasekhar-Clogston limit \cite{kujawa}, \cite{chandrasekhar}.
%However, the existence of RT does not influence the existence of the region with  $\rho_s<0$, i.e. even if there is no RT in the phase diagrams with the Hartree term, there is a region with $\rho_s<0$ in the weak coupling limit. 
%Above the dotted red line there is a spin-polarized region which has a gapless spectrum for the majority spin species. 
%This observation points that there exists an instability of the homogeneous superconducting state and, in turn, suggests the existence of a FFLO state in the weak coupling regime. 

%At $T=0$ and $h\neq 0$, we have the following sequences of transitions (see Fig. 1): SC$_0$ $\rightarrow$ NO or SC$_0$ $\rightarrow$ PS $\rightarrow$ NO. When the temperature increases, the character of the transition changes from first to second order and the phase separation lines meet with the 2$^{nd}$ order phase transition line at TCP. 
\subsection{\textbf{(iii)} $\boldsymbol{r\neq 1}$, $\boldsymbol{h\neq 0}$, $\boldsymbol{T=0}$} The BCS-{LP} crossover diagrams in the presence of a Zeeman magnetic field for $r\neq 1$ exhibit a novel behavior. As opposed to the $r=1$ case, for strong attraction, SC$_M$ occurs at $T=0$ (Fig. \ref{fig3}). In general, these types of solutions (Sarma-type with $\Delta (h)$) appear (for $r{>} 1$) when $h>(\frac{r-1}{r+1})\bar{\mu}+2\Delta \frac{{\sqrt{r}}}{r+1}$ (on the BCS side) or when  $h>{\sqrt{(\bar{\mu}-\epsilon_0)^2+|\Delta |^2}-D\frac{r-1}{r+1}}$ (on the {LP} side). The $SC_M$ phase is unstable in the weak coupling regime at $T=0$, but can be stable in the intermediate and strong coupling LP limit. Deep in the {LP} limit, unpaired spin down fermions do not exist. Hence, the SC$_M$ phase is the superfluid state of coexisting LP's (hard-core bosons) and single-species fermions, with the latter responsible for finite polarization (magnetization) and the 
gapless excitations characteristic for this state of Bose-Fermi mixture. 

The structure of the ground state diagrams in Fig. \ref{fig3} is different from those shown in Fig. \ref{fig1}, where one has only a first order phase transition from pure SC$_0$ to the NO phase in the ($\mu -h$) plane. In addition, there exist critical values of $|U|$ ($|U_c|^{SC_M}$ -- red points in the diagrams), for which the SC$_M$ state becomes stable, instead of PS. However, one should mention that there is a critical value of $r$, for which SC$_M$ is stable. Fig. \ref{fig3}(a) shows the ground state ($r-E_b/E_F$) phase diagram for fixed $n=0.01$ and $h=0$. The SC$_M$ state does not appear stable up to $r\approx 5$ in the diagram with the Hartree term and also up to $r\approx 4.2$ in the diagram without the Hartree term. The presence of such a term restricts the range of occurrence of SC$_M$, except for a very dilute limit. 

The diagrams ($h-|U|$) for higher filling ($n=0.2$) and fixed $r$ are presented in Fig. \ref{fig3}(b), (c). For higher $n$, the region of SC$_M$ is narrowing. The SC$_M$ state is unstable even at $r=5$, in the diagram without the Hartree term. The transition from SC$_M$ to FP can be accomplished in two ways: through PS-III (SC$_M$+FP) or through a 2$^{nd}$ order phase transition. The character of this transition changes with increasing $|U|$. In the very strong coupling limit, PS is more stable than the SC$_M$ phase. Hence, we also find two TCP in these diagrams (Fig. \ref{fig3}(c)). However, in the very dilute limit ($n\rightarrow 0$) there is only one TCP in the ($h-|U|$) diagram. Therefore, we can distinguish the following sequences of transitions: SC$_0$ $\rightarrow$ PP (FP) or SC$_0$ $\rightarrow$ SC$_M$ $\rightarrow$ FP. 

In fact, SC$_0$ $\rightarrow$ SC$_M$ is a topological quantum phase transition (Lifshitz type). Across this transition there is a cusp in the order parameter and the chemical potential vs. magnetic field plots. There is also a change in the electronic structure. In the SC$_0$ phase, there is no FS, but in the SC$_M$ state, there is one FS for excess fermions. It is worth mentioning that the value of $|U|$ for which $\bar{\mu}$ reaches the lower band edge does not depend on the mass imbalance in the SC$_0$ state. The BP-2 phase in $d=2$ is unstable, even for large mass ratio. If $r\neq 1$, the  symmetry with respect to $h=0$ is broken. However, this symmetry is restored upon replacement $r\to r^{-1}$. 

%\noindent\textbf{(iv)}
 \subsection{\textbf{(iv)} $\boldsymbol{r\neq 1}$, $\boldsymbol{h=0}$, $\boldsymbol{T\neq 0}$}  Fig. \ref{fig4} shows the ($r-|U|$) ground state diagram (a) and also the ($T-|U|$) phase diagrams, for fixed $h=0$, $r=4$ (b) and $r=7$ (c). At $T=0$, $r\neq 1$, we have the following sequences of transitions with increasing $|U|$ (see Fig. \ref{fig4}(a)): FP $\rightarrow$ SC$_M$ $\rightarrow$ SC$_0$ (for higher values of $r$) or FP (PP) $\rightarrow$ PS $\rightarrow$ SC$_0$ (for lower values of $r$). 
Because of the occurrence of the $SC_M$ state at $T=0$ for higher $r$, this phase persists to non-zero temperatures (as shown in Fig. \ref{fig4}(c), $r=7$). However, if $SC_M$ is unstable at $T=0$ for lower $r$ (Fig. \ref{fig4}(b)), the gapless region can still occur at some temperatures (with one FS in the strong coupling). %The solid lines (2$^{nd}$ order phase transition lines) and PS regions are determined within the BCS approximation. 
The system is a quasi superconductor below $T_c^{KT}$. Apart from unpolarized qSC$_0$ state, qSC$_M$ occurs which can be termed a spin polarized KT superfluid (Fig. \ref{fig4}(c)).  Above $T_c^{KT}$ we have an extended region of incoherent pairs which is bounded from above by the pair breaking temperature. In the strong coupling limit,  $T_c^{KT}$ does not depend on magnetic field, but it depends on mass imbalance and takes the form: $k_{B}T_c^{KT}=2\pi\frac{r}{(1+r)^2}\frac{t^2}{|U|}n(2-n)$ ($r>0$). In that limit only LP's exist and the system is equivalent to that of hard-core Bose gas on a lattice, described by the Hamiltonian (\ref{pseudospin}).Because of the existence of the SC$_M$ state at $T=0$ in the $r\neq1$ case, the range of occurrence of a spin polarized KT superfluid is much larger than for $r=1$ (see inset in Fig. \ref{fig4}(c)).
%At $r=1$ However, if $r\neq1$ this phase exists at fixed polarization. 

\section{Conclusions} We have investigated the evolution from the weak coupling to the strong coupling limit of tightly bound local pairs with increasing attraction for $d=2$, within the spin-polarized AHM. If the number of particles is fixed and $n\neq 1$, one obtains two critical Zeeman magnetic fields, which limit PS of the $SC_0$ (or SC$_M$) and the NO states. The occurrence of the BP-1 phase depends on the lattice structure, i.e. if $r=1$, SC$_M$ is unstable for $d=2$ but it can be realized for $d=3$ lattices \cite{kujawa2}. However, in the AHM the very existence of the BP-1 phase is restricted to low fillings. The Hartree term, usually promoting ferromagnetism in the Stoner model ($U>0$), here ($U<0$) strongly competes with superconductivity. Thus, such a term restricts the SC$_M$ state to lower densities. However, the mass imbalance can change this behaviour even for $d=2$ due to spin polarization stemming from the kinetic energy term. In this way, $SC_M$ can be realized in $d=2$ for the intermediate 
and strong coupling regimes, but there is a critical value of the mass ratio for which SC$_M$ is stable, at finite fixed $n$. In other words, the combination of mass and population imbalance can stabilize BP-1 phase in 2D, on the BEC side of crossover. We also determined the critical value of {$n$} {above} which {SC} {and} {CO} {can form} {PS state} at $h=0$ and $r\neq 1$. We have found that the BP-2 state is unstable in the whole range of parameters, in the $d=2$ one-band spin-polarized AHM. Nevertheless, one can suppose that the Liu-Wilczek (BP-2) phase can be realized within the two-band model. The TCP's were found in the $(h-|U|)$ diagrams at $r\neq 1$ and $T=0$. 
We have also extended {the analysis of the crossover} to finite temperatures in $d=2$ by invoking the KT scenario. The KT transition temperatures are definitely lower than the ones determined in the BCS scheme. Moreover, spin polarization has a strong destroying influence on the KT superfluid state at $r=1$ and allows this phase in the weak coupling regime, in agreement with the results for the continuum case \cite{tempere}. In the strong coupling limit, $T_c^{KT}$ does not depend on magnetic field (below $h_{c1}$) and for $k_BT<<|U|$ only unbroken LP's exist, which can form unpolarized qSC below $T_c^{KT}$ or stay phase disordered. However, if $r\neq 1$, a spin polarized KT superfluid state can be stable even in the intermediate and strong coupling region. In this work we have not considered nonhomogeneous FFLO states which are possible in weak to intermediate attraction range, albeit much more susceptible to phase fluctuations at finite $T$ in 2D system \cite{Shimahara}. 
%An extended analysis taking into consideration a competition between the superconducting phases, CO and unconventional spin density wave (SDW) within the spin polarized Hubbard model is underway.
 
\begin{acknowledgments}
{A. K.-C. acknowledges financial support under grant No. N N202 030540 (MSHE-Poland).} 
We thank R. W. Chhajlany and S. Robaszkiewicz for carefully reading manuscript and valuable comments. 
\end{acknowledgments}

\end{document}